\documentclass[twocolumn,preprintnumbers,amsmath,amssymb,pra]{revtex4}
\usepackage{txfonts}
\usepackage{graphicx,hyperref}
\usepackage{dcolumn}
\usepackage{bm}

\usepackage{color}

 \allowdisplaybreaks
\begin{document}


\title{Phase control of localization in the nonlinear two-mode system from harmonic mixing driving: Perturbative analysis and symmetry consideration}

\author{Xianchao Le$^{1}$}
\author{Zhao-Yun Zeng$^{2}$}
\author{Baiyuan Yang$^{2}$}
\author{Yunrong Luo$^{3}$}
\author{Jinpeng Xiao$^{2}$}
\author{Lei Li$^{2}$}
\author{Lisheng Wang$^{2}$}
\author{Yajiang Chen$^{2}$}
\author{Ai-Xi Chen$^{1}$}
\author{Xiaobing Luo$^{1,2}$}
\altaffiliation{Corresponding author: xiaobingluo2013@aliyun.com}
\affiliation{$^{1}$Department of Physics, Zhejiang Sci-Tech University, Hangzhou, 310018, China}
\affiliation{$^{2}$School of Mathematics and Physics, Jinggangshan University, Ji'an 343009, China}
\affiliation{$^{3}$ Department of Physics and Key Laboratory for Matter Microstructure and Function of Hunan Province, and Key Laboratory of
Low-dimensional Quantum Structures and Quantum Control of Ministry
of Education, Hunan Normal University, Changsha 410081, China}

\date{\today}

\begin{abstract}
In this paper, we present a rigorous analysis of symmetry and underlying physics of the nonlinear two-mode system
driven by a harmonic mixing field, by means of multiple scale asymptotic analysis method.  The effective description in the framework of the second-order perturbative theory provides an accurate picture for understanding the Floquet eigenspectrum and dynamical features of the nonlinear two-mode system, showing full agreement with the prediction of symmetry considerations. We find that two types of symmetries play significant role in the dynamical features of this model, the mechanism behind which can be interpreted in terms of the effective description.    The results are of relevance for the phase control of the atomic localization in Bose-Einstein condensates or switch of the optical signals in nonlinear mediums.\\

\end{abstract}

\maketitle
\section{introduction}
Nonlinear two-mode model is one prototypical example to investigate the fundamental
quantum effects and nonlinear tunneling dynamics. The existence of nonlinearity is ubiquitous in diverse branches of science and its physical origins
include a mean-field treatment of the interactions between coherent atoms\cite{Pethick2022}, nonlinear Kerr effects in optical fibers\cite{Agrawal}, and possible modifications of quantum mechanics
on the fundamental level\cite{Weinberg1989}. In reality, the nonlinear two-mode model can be applied to describe a wide variety of physical systems, such as two coupled optical waveguides with Kerr nonlinearity\cite{Jensen1982}, Bose-Einstein
condensates (BECs) in a double-well potential\cite{Smerzi1997}, among others. As is well known, the presence of nonlinearity gives rise to a number of new quantum natures such as macroscopic quantum self-trapping
(MQST)\cite{Smerzi1997, Milburn1997, Albiez2005} and breakdown of quantum
adiabaticity\cite{Wu2000, Liu2002, Liu2003}, which presents both challenges and opportunities for existing theories
in linear systems.

In recent years, Floquet engineering, i.e., coherent control via periodic driving, has offered a versatile method for realization of  new phases not accessible in equilibrium systems\cite{Bukov2015, Eckardt2017, Silveri2017, An2021}, because it adds time-periodicity as a novel control dimension to quantum systems. Nonlinear two-mode model under periodic driving presents the paradigm of this hot topic and thus is attracting more and more interest, due to the fact that gaining insight from the combined effects
of periodic driving and nonlinearity on the quantum tunneling through a barrier may contribute to the possibility of utilizing two-state systems as the basic building blocks of
quantum-based devices. In
periodically driven quantum systems, it is convenient to analyze the dynamics
in terms of the so-called Floquet states and quasi-energies. When nonlinearity is introduced, it is necessary
to extend the conventional Floquet states to nonlinear
Floquet states\cite{Holthaus12001, Holthaus22001, Luo2007, Luo2008, Molina2008}. Nowadays considerable efforts have been devoted to study the periodically driven nonlinear two-state systems in different ways, i.e., by
employing an effective Hamiltonian
description\cite{Wang2006, Zhang2008}, numerically computing the nonlinear
Floquet quasienergy spectrum\cite{Luo2007, Luo2008, Molina2008, Molina2008R, Lyu2020}, and  constructing the exact analytical nonlinear Floquet solutions\cite{Xie12007, Xie22007, Yang2016} as well. Rich dynamical behaviors have been uncovered, such as the emergence of Hamiltonian
chaos\cite{Abdullaev2000, Lee2001, Hai2002, Weiss2008, Jiang2014}, photon-assisted
tunneling\cite{Eckardt2005, Watanabe2010}, coherent control of self-trapping\cite{Holthaus12001, Holthaus22001, Wang2006, Xie12007}, and so
on.

On the other side, previous works have clearly identified that the symmetries of the time-periodic
Hamiltonian play crucial role in the current rectification
phenomenon or the ratchet effect in the driven periodic potential\cite{Reimann2002, Hanggi2009}. It has been shown that in order to
achieve directed
(ratchet) transport,
relevant symmetries have to be broken\cite{Flach2000, Denisov2007}.
According to the Curie's principle,
a certain phenomenon always occurs unless it is ruled out by
symmetries\cite{Reimann2002}.
The discussion of  space-time symmetries was also exemplified nicely by the
two-state dynamics\cite{Kierig2008}. It has been noted that a broken space-time symmetry leads in general to a driving-induced unbalanced Floquet states with unequal population of two modes. In addition, a prominent quantum effect called coherent destruction of tunneling (CDT)\cite{Grossmann1991}, upon
the occurrence of which the quantum tunneling effects can be completely suppressed, has been shown to be connected to the degeneracy
of the quasienergies and the so-called generalized parity symmetry (that is, the Hamiltonian
is invariant under a spatial parity transformation plus a time
shift by half a driving period)\cite{Kierig2008}. An important question now arises as to how the symmetry pictures are modified when
the two-state system is subject to nonlinearity. Despite a few numerical
studies available in such a problem, the connection between  symmetry and dynamical properties of the driven nonlinear two-mode model is still not very clear, which awaits more rigorous and
elaborate analytical results.

In the present work, we shall explore the symmetry and the underlying physics of the nonlinear two-state system
exposed to a harmonic mixing field, in a
rigorous way by means of multiple-scale asymptotic analysis method.
By pushing the multiple-time-scale asymptotic analysis up to the second order, it is shown that apart from renormalization of the tunneling parameter,
the harmonic mixing driving  may also induce an effective static dc-bias between two modes whose amplitude and sign depend on the phase shift between two harmonics.
The effective time-independent Hamiltonian  obtained in the framework of the second-order perturbative theory is found to be successful in capturing the fine structure of the
quasienergy spectrum with different time-space symmetries, which confirms the
predictions of the symmetry considerations.
The analytical results give a clear explanation for why the phase shift between the two-harmonics components of driving field can play the role of control
parameter for the amplitude and sign of the population imbalance of nonlinear Floquet states.

\section{Model and symmetry}
We consider a simple, yet non-trivial periodically driven nonlinear two-mode system consisting of two basis states
$|1\rangle$ and $|2\rangle$, whose dynamics  is described by
\begin{align}\label{Dnls1}
  i\frac{dc_1}{dt}= &-\frac{v}{2}c_2+\frac{S(t)}{2}c_1-\chi|c_1|^2c_1\nonumber \\
  i\frac{dc_2}{dt}= &-\frac{v}{2}c_1-\frac{S(t)}{2}c_2-\chi|c_2|^2c_2,
\end{align}
where $v$ denotes the tunneling rate constant, $\chi$ is the nonlinearity strength, $c_{1,2}$ are the quantum
probability amplitudes on the two basis states $|1\rangle$ and $|2\rangle$, and $S(t)$ is an external periodic field of zero mean,
$S(t+T)=S(t)$.

Before proceeding to the analysis of dynamics of the system \eqref{Dnls1}, it is instructive to identify the symmetry property of the model
equation. Like its linear counterpart, the driven nonlinear two-mode system also admits solutions in the form of Floquet
states $\mathbf{c}(t)=\tilde{\mathbf{c}}(t)e^{-i\varepsilon t}$, where $\mathbf{c}(t)=[c_1(t),c_2(t)]^T$ (hereafter the superscript $T$ stands for the
 transpose), $\varepsilon$ is the quasienergy,
 and $\mathbf{\tilde{c}}(t)=[\tilde{c}_1(t),\tilde{c}_2(t)]^T$  inherits
the period of the driving  and is called the Floquet eigenstate. Substituting the Floquet solution  into Eq.~\eqref{Dnls1}, we obtain
the following eigenvalue equation
 \begin{align}\label{eigen eq}
\mathcal{H}\mathbf{\tilde{c}}(t)=\varepsilon\mathbf{\tilde{c}}(t),~~\mathcal{H}:=H(t)-i\partial_t,
 \end{align}
with time-periodic Hamiltonian
 corresponding to the system \eqref{Dnls1}, i.e.,
 \begin{equation}\label{Ht}
  H(t)=\left(
    \begin{array}{cc}
      \frac{S(t)}{2}-\chi |\tilde{c}_1(t)|^2 & -\frac{v}{2} \\
      -\frac{v}{2} & -\frac{S(t)}{2}-\chi |\tilde{c}_2(t)|^2 \\
    \end{array}
  \right),
\end{equation}
where the operator $\mathcal{H}$ is the so-called Floquet Hamiltonian defined in the extended Hilbert space.

For the linear ($\chi=0$) case, we should review the three relevant symmetries
of \eqref{eigen eq} below. If $S(t)$ is shift symmetric
$S(t)=-S(t+T/2)$, then $\mathcal{H}$ is invariant under the generalized parity symmetry
\begin{equation}\label{SA}
S_{\rm{GP}}: |1\rangle\leftrightarrow |2\rangle,~~t\rightarrow t+\frac{T}{2},
\end{equation}
which consists of a spatial parity transformation plus a time shift by half a driving
period.

If $S(t)$ is antisymmetric under $t\rightarrow-t+2t_0$ inversion, $S(t+t_0)=-S(-t+t_0)$ at some appropriate points $t_0$,
the Floquet Hamiltonian is invariant under the following symmetry
\begin{equation}\label{SB}
S_{\rm{PT}}: |1\rangle\leftrightarrow |2\rangle,~~t\rightarrow -t+2t_0, i\rightarrow-i,
\end{equation}
which is equivalent to parity-time symmetry. The transformation $S_{\rm{PT}}$ represents the combined  parity
and time reversal operations.

Furthermore, if $S(t)$ possesses the symmetry $S(t+t_0) =S(-t+t_0)$, then the Floquet Hamiltonian $\mathcal{H}$ is
time-reversal invariant under
\begin{equation}\label{SB}
S_{\rm{T}}: ~~t\rightarrow -t+2t_0, i\rightarrow-i.
\end{equation}

To measure the localization
properties of a Floquet mode, we use the time-averaged expectation values of the
Pauli matrix $\sigma_z$:
\begin{equation}\label{expectation}
\langle\langle\sigma_z\rangle\rangle=\frac{1}{T}\int_0^{T}dt\mathbf{\tilde{c}}^{\dag}(t)\sigma_z\mathbf{\tilde{c}}(t).
\end{equation}
This time-averaged expectation value $\langle\langle\sigma_z\rangle\rangle$ would vanish with a perfectly delocalized Floquet state (balanced Floquet state).

In the linear limit ($\chi=0$), it can be readily verified that the system \eqref{Dnls1}  has two Floquet states
with zero time-averaged population imbalances, namely $\langle\langle\sigma_z\rangle\rangle=0$, whenever the symmetries $S_{\rm{GP}}$ and/or $S_{\rm{PT}}$
are realized.

Next we will show what consequences arise for the above mentioned symmetries when the nonlinearity is imposed. Consider a harmonic mixing (two-frequency) driving
\begin{equation}\label{expectation}
S(t)=-A[\sin\omega t+f\sin(2\omega t+\phi)],
\end{equation}
which has two components of frequencies $\omega$
and $2\omega$ with  phase shift $\phi$. Obviously, in the presence of both harmonics ($A\neq 0,~f\neq0 $), the generalized
parity symmetry $S_{\rm{GP}}$ is always violated, independently of the value of the
phase shift $\phi$. Note that the antisymmetry $S(t)=-S(-t)$ is preserved for
$\phi=n\pi$ with $n$ integer, and the time-reversal symmetry $S(t_0+t)=S(t_0-t)$ is preserved for $\phi=n\pi+\pi/2$.

The relationship between the antisymmetry $S(t)=-S(-t)$ and dynamical properties can be readily analyzed by the symmetry argument.    We implement $S_{\rm{PT}}$ on the nonlinear model Eq.~\eqref{eigen eq}, where $S_{\rm{PT}}$ includes the
combined parity and time reversal operations. As the first step, time-reversal transformation (which changes a complex number to its
complex conjugate and turns $t$ into $-t$) convert Eq.~\eqref{eigen eq} into
\begin{align}\label{PT equ}
  [\widetilde{H}-i\partial_t]\left(
             \begin{array}{c}
              \tilde{c}_1^*(-t) \\
              \tilde{c}_2^*(-t) \\
             \end{array}
           \right)=\varepsilon\left(
             \begin{array}{c}
              \tilde{c}_1^*(-t) \\
              \tilde{c}_2^*(-t) \\
             \end{array}
           \right),
\end{align}
with
\begin{align}\label{PT equ H}
\widetilde{H}=\left(
    \begin{array}{cc}
      \frac{S(-t)}{2}-\chi |\tilde{c}_1^*(-t)|^2 & -\frac{v}{2} \\
      -\frac{v}{2} & -\frac{S(-t)}{2}-\chi |\tilde{c}_2^*(-t)|^2 \\
    \end{array}
  \right).
  \end{align}
The second step is to act with the parity transformation (permutation of the two indices 1 and 2) on Eq.~\eqref{PT equ}, from which we can observe
\begin{equation}\label{PT equ2}
\mathcal{H'}\left(
             \begin{array}{c}
             \tilde{c}_2^*(-t) \\
              \tilde{c}_1^*(-t) \\
             \end{array}
           \right)=\varepsilon\left(
             \begin{array}{c}
              \tilde{c}_2^*(-t) \\
              \tilde{c}_1^*(-t) \\
             \end{array}
           \right)
\end{equation}
with
\begin{equation}\label{PT H2}
\mathcal{H'}=\left(
    \begin{array}{cc}
      -\frac{S(-t)}{2}-\chi |\tilde{c}_2^*(-t)|^2 & -\frac{v}{2} \\
      -\frac{v}{2} & \frac{S(-t)}{2}-\chi |\tilde{c}_1^*(-t)|^2 \\
    \end{array}
  \right)-i\partial_t.
\end{equation}

By comparing  Eq.~\eqref{PT H2} with its original equation \eqref{eigen eq}, we find that for a nonlinear system under the action of antisymmetric driving [$S(t)=-S(-t)$], the
$S_{\rm{PT}}$ operation leaves  the Floquet Hamiltonian invariant, namely $\mathcal{H'}=\mathcal{H}$, provided that
the following constraint,
\begin{equation}\label{constraints}
\chi|\tilde{c}_2(-t)|=\chi|\tilde{c}_1(t)|,
\end{equation}
is satisfied.
If the Floquet Hamiltonian is invariant under $S_{\rm{PT}}$, we have
\begin{equation}\label{Relat twosolu}
 S_{\rm{PT}}\left(
             \begin{array}{c}
              \tilde{c}_1(t) \\
              \tilde{c}_2(t) \\
             \end{array}
           \right)= \left(
             \begin{array}{c}
              \tilde{c}_2^*(-t) \\
              \tilde{c}_1^*(-t) \\
             \end{array}
           \right)=\left(
             \begin{array}{c}
              \tilde{c}_1(t) \\
              \tilde{c}_2(t) \\
             \end{array}
           \right)e^{i\varphi}.
\end{equation}
Here the Floquet states are defined up to an arbitrary phase $\varphi$. Physically, Eq.~\eqref{Relat twosolu} implies that the Floquet Hamiltonian operator $\mathcal{H}$
and the $S_{\rm{PT}}$ operator share the same eigenmode. Once Eq.~\eqref{Relat twosolu} is satisfied, the constraint \eqref{constraints} is satisfied automatically and vice verse.

From \eqref{Relat twosolu}, it is easy to prove
\begin{align}
  \frac{1}{T}\int_0^T|\tilde{c}_2(t)|^2dt=\frac{1}{T}\int_{0}^{T}|\tilde{c}_1(t)|^2dt,
\end{align}
which means $\langle\langle\sigma_z\rangle\rangle=\frac{1}{T}\int_0^T[|c_1(t)|^2-|c_2(t)|^2]dt=0$, representing balanced  Floquet states with zero averaged population imbalances. Apparently, whether  nonlinearity presents or not, the balanced Floquet states exist inevitably for the antisymmetric driving. Nevertheless, if the antisymmetry $S(t)=-S(-t)$ is violated, the balanced Floquet states will disappear since the Floquet Hamiltonian changes under the action of operation $S_{\rm{PT}}$, and all the Floquet states  will acquire some nonzero population
imbalances.

There are exceptions for the nonlinear case.
If the antisymmetry  $S(t)=-S(-t)$ holds, while $|\tilde{c}_1(t)|^2=|\tilde{c}_2(-t)|^2$ does not hold, the Floquet Hamiltonian in \eqref{eigen eq} is not invariant under $S_{\rm{PT}}$ symmetry operation. This situation will lead to the surprising result that two doubly-degenerate unbalanced nonlinear Floquet
states emerge, as described as follows. Due to $S(t)=-S(-t)$, from \eqref{eigen eq} and \eqref{PT H2} it follows that
the system admits two independent Floquet solutions
$|\psi_1\rangle=[\tilde{c}_1(t),\tilde{c}_2(t)]^T$ and $|\psi_2\rangle=[\tilde{c}_2^*(-t),\tilde{c}_1^*(-t)]^T$ corresponding to the same quasienergy $\varepsilon$.
In this case, because of  $|\tilde{c}_1(t)|^2\neq|\tilde{c}_2(-t)|^2$, it is easy to see that
\begin{align}
\langle\langle\sigma_z\rangle\rangle=\frac{1}{T}\int_0^T[|c_1(t)|^2-|c_2(t)|^2]dt\neq 0,\\
\frac{1}{T}\int_0^T\langle\psi_1|\sigma_z|\psi_1\rangle dt=-\frac{1}{T}\int_0^T\langle\psi_2|\sigma_z|\psi_2\rangle dt,
\end{align}
which denotes the emergence of two doubly-degenerate unbalanced (localized) nonlinear Floquet
states, with exactly opposite time-averaged population imbalance.
Apparently, these two degenerate Floquet states exist only when the nonlinear term does not vanish, so
that they have no linear counterparts. Also note that the degeneracy for the nonlinear Floquet states can be lifted by breaking of the antisymmetry $S(t)=-S(-t)$.

\section{Quasienergies and Floquet states}
In this section, we shall numerically compute the nonlinear Floquet
states and corresponding quasienergies by following the strategy developed in Refs.~\cite{Luo2007, Luo2008}.
In this strategy, we expand $\tilde{\mathbf{c}}(t)$ as well as
the time-periodic modulation $S(t)$ into Fourier series with $2N +1$ modes: $\tilde{c}_1\left( t \right) =\sum_{n=-N}^N{a_n}e^{in\omega t}$,  $\tilde{c}_2\left( t \right) =\sum_{n=-N}^N{b_n}e^{in\omega t}$, and
$S\left( t \right) =\sum_{m=-N}^N{p_m}e^{im\omega t},~ p_m=\frac{1}{T}\int_0^T{S\left( t \right)}e^{-im\omega t}dt$. Substituting these series into Eq.~\eqref{eigen eq} yields
\begin{align}\label{egienI}
&\sum_{n,m}{a_n}p_me^{i\left( m+n \right) \omega t}-\chi \sum_{n,m,m'}{a_m{a^*}_{m'}a_{n}}e^{i\left( m+n-m' \right) \omega t}-\frac{v}{2}\sum_n{b_n}e^{in\omega t}\nonumber\\
&+\sum_n{n\omega a_n}e^{in\omega t}=\varepsilon \sum_n{a_n}e^{in\omega t},\nonumber\\
&-\sum_{n,m}{b_n}p_me^{i\left( m+n \right) \omega t}-\chi \sum_{n,m,m'}{b_m{b^*}_{m'}b_{n}}e^{i\left( m+n-m' \right) \omega t}-\frac{v}{2}\sum_n{a_n}e^{in\omega t}\nonumber\\
&+\sum_n{n\omega b_n}e^{in\omega t}=\varepsilon \sum_n{b_n}e^{in\omega t}.
\end{align}

Multiplying the above equations by $e^{-ij\omega t}$, and integrating them over one driving period, one gets the following eigenvalue equation,
\begin{align}\label{egienII}
&-\frac{A}{4i}\left( a_{j-1}-a_{j+1}+fe^{i\phi}a_{j-2}-fe^{-i\phi}a_{j+2} \right) -\chi \sum_{m,m'}{a_m{a^*}_{m'}a_{j+m'-m}}\nonumber\\&-\frac{\nu}{2}b_j+j\omega a_j=\varepsilon a_j,\nonumber\\
&\frac{A}{4i}\left( b_{j-1}-b_{j+1}+fe^{i\phi}b_{j-2}-fe^{-i\phi}b_{j+2} \right) -\chi \sum_{m,m'}{b_m{b^*}_{m'}b_{j+m'-m}}\nonumber\\&-\frac{\nu}{2}a_j+j\omega b_j=\varepsilon b_j.
\end{align}

Finally, the nonlinear Floquet
states and corresponding quasienergies can be found by solving  numerically the  eigenvalue equation \eqref{egienII} in a
self-consistent manner. In numerical calculations, the Fourier terms of orders
higher than a cut-off order $N$ is neglected when convergence is achieved. The above procedure involves conservation of the
norm, i.e.,
\begin{align}\label{norm}
\sum_n |a_n|^2+\sum_n |b_n|^2=1.
\end{align}

In analogy to quasimomenta
in the spatially periodic crystal, the quasienergy spectrum repeats itself
periodically on the energy-axis, thus possessing Brillouin zonelike
structure, the width
of one zone being $\hbar\omega$ ($\hbar=1$). In the following, we restrict ourselves to states with quasienergies in one Brillouin zone $(-\omega/2, \omega/2]$.

\begin{figure}[htbp]
\center
\includegraphics[width=8cm]{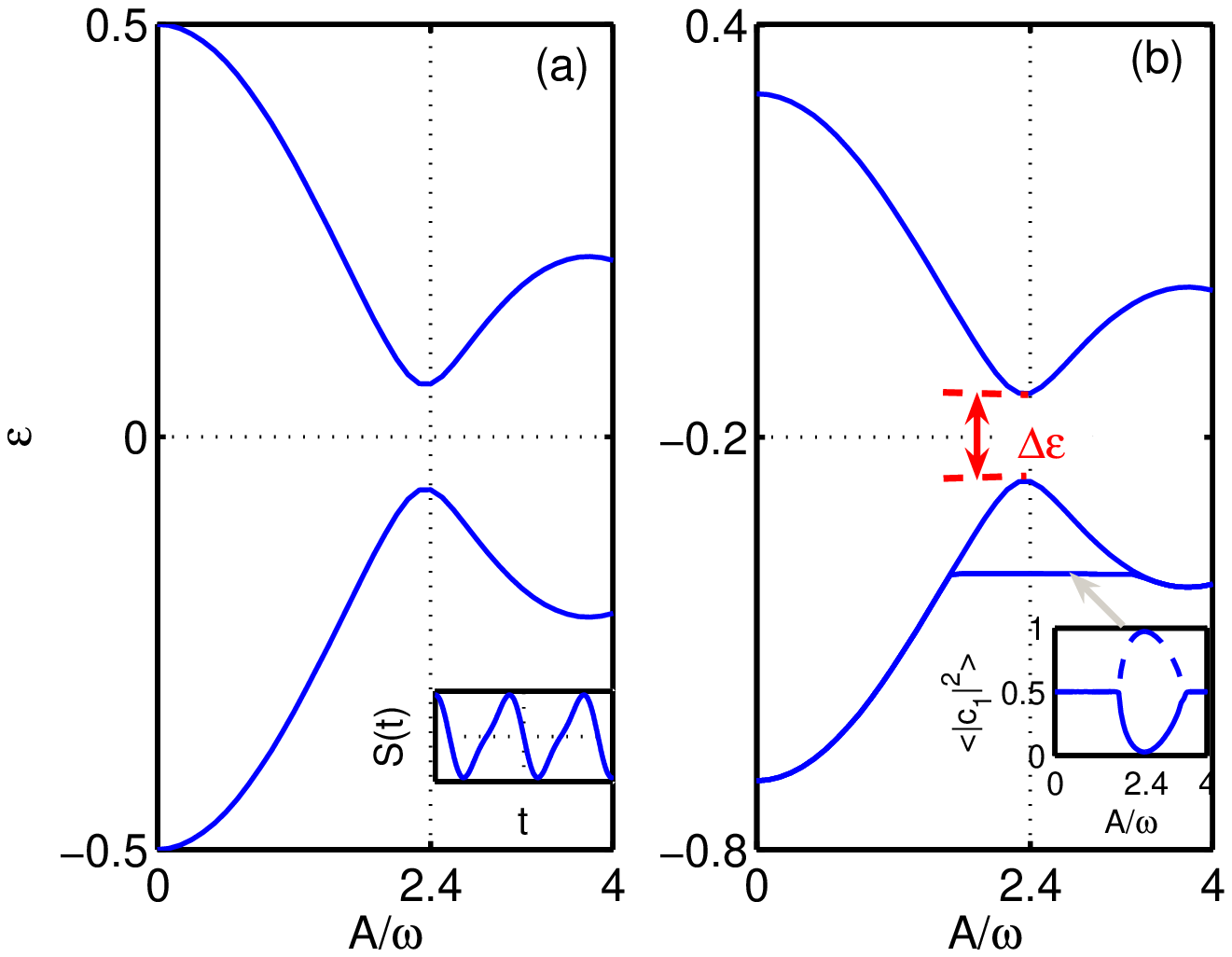}
\includegraphics[width=8cm]{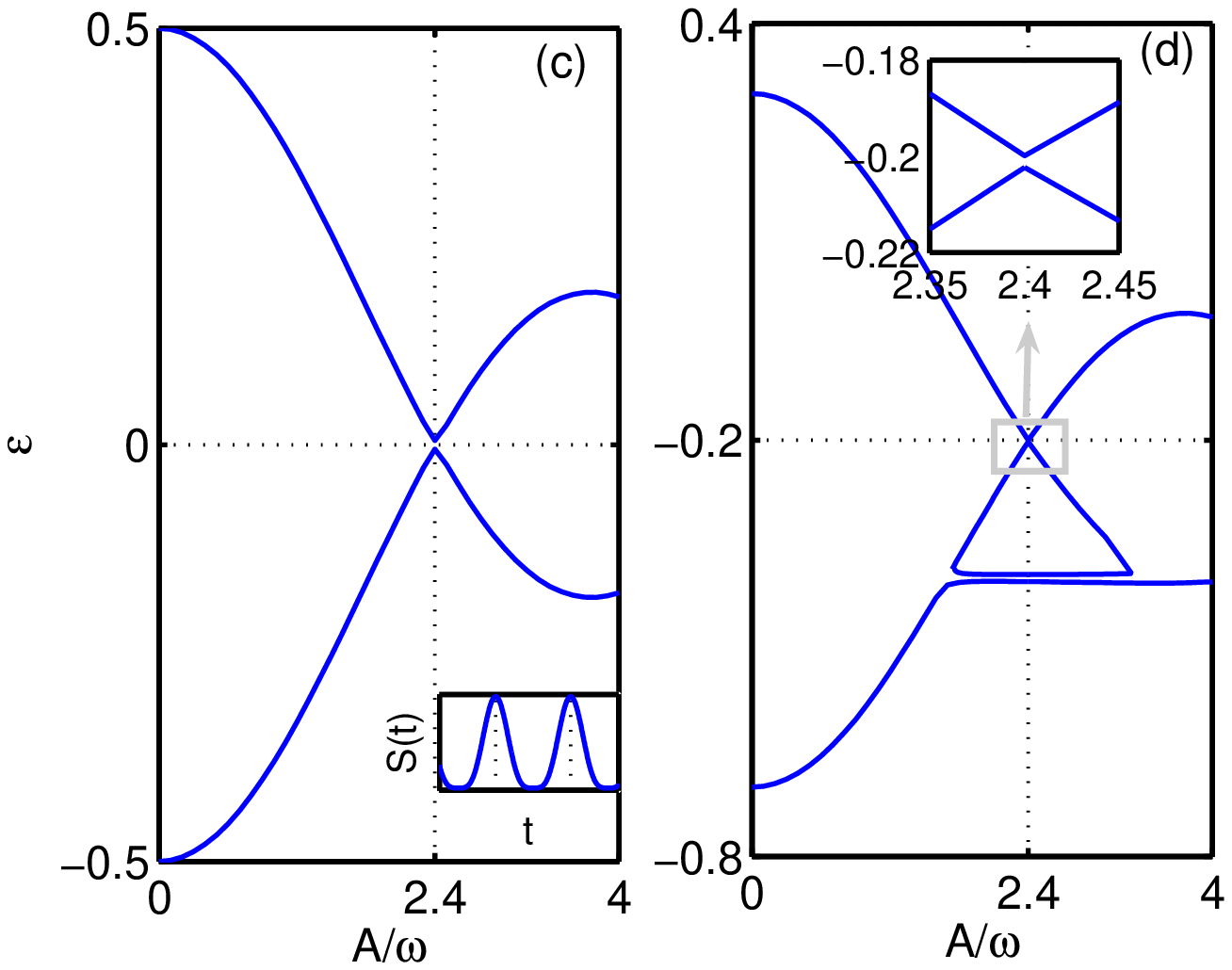}
\caption{(color online) The Floquet state quasienergies versus the driving parameter $A/\omega$. [(a), (b)] the case of time-reversal antisymmetric driving [see the profile of $S(t)$ at $\phi=0$, in the bottom-right inset of panel (a)]. [(c), (d)] the case of time-reversal symmetric driving [see the shape of $S(t)$ at $\phi=\pi/2$ in
the inset of panel (c)]. The left column [(a), (c)] is for the linear case $\chi=0$, while right column [(b), (d)] is for the nonlinear case $\chi=0.4$. The inset of panel (b) shows the  time-averaged population of mode 1 for every Floquet state in the
lowest quasienergy level. Inset of panel (b) is the enlarged view of the closest approach of two normal Floquet states at $A/\omega=2.4$. Other parameters: $f=1/4, \omega=10, v=1$. }\label{fig1}
\end{figure}
Our numerical results of quasienergies are plotted in Fig.~\ref{fig1}. Figs.~\ref{fig1} (a)-(b) illustrate the quasienergies with an antisymmetric  but a sawtooth driving [namely, $S(t)=-S(-t)$ with $\phi=0$, see bottom-right inset in Fig.~\ref{fig1} (a)], and Figs.~\ref{fig1} (c)-(d) illustrate the ones with a symmetric  driving [namely, $S(t+t_0)=S(-t+t_0)$ with $\phi=\pi/2$, see bottom-right inset in Fig.~\ref{fig1} (c)].
It is clear from Fig.~\ref{fig1} that there are two
quasienergies at a given value of $A/\omega$ for the linear case (see the left column), while in the presence of nonlinearity several new states are emerging within certain range of  $A/\omega$  due to bifurcation (see the right column). As the nonlinearity sets in, the antisymmetric driving case shows a pitchfork bifurcation with appearance of the additional quasienergy level that is absent in the linear case and lies in the lowest branch. As seen in the inset of Fig.~\ref{fig1} (b), the additional quasienergy is in fact doubly degenerate and corresponds
to two different Floquet states with exactly opposite
nonzero population imbalance, which can be witnessed by calculation of the cycle-averaged population, $\langle |c_1|^2\rangle=\frac{1}{T}\int_0^T|c_1|^2 dt$, for the given Floquet state. By contrast, for the symmetric driving [see Fig.~\ref{fig1} (d)], the twofold degeneracy for the lowest quasienergy is lifted and the quasienergy splits into two quasi-degenerate (nearly coincident) quasienergy levels. Among the two  quasi-degenerate quasienergy levels that have no linear equivalent, the lower one shows strict continuation from the undriven limit, while the other emerges
through the saddle-node bifurcation upon the smooth change of the driving parameter $A/\omega$.
It should be noted that there is no threshold value of nonlinearity for
level bifurcation (forming the well-known triangular structure) to appear  for the two-mode system under symmetric driving, which is the same as the purely sinusoidal driving case\cite{Luo2007, Luo2008}. However, for the antisymmetric driving case, our numerical results, which are not listed here, reveal that level bifurcation occurs only above a certain critical value of nonlinearity, the reason of which  will be explained later. In addition to the new quasienergies emerging from bifurcation, there  also exists two normal Floquet states which survive for vanishing nonlinearity and  thus have linear counterparts. It is clear that the two normal Floquet states make closest approach at $A/\omega=2.4$. A significant difference between the antisymmetric and symmetric case is that for the former, there exists a large gap between the two normal Floquet state levels [characterized by the minimal level spacing between  the two Floquet states, see $\Delta\varepsilon$ as labeled in Fig.~\ref{fig1} (b)], while for the latter, there is a nearly vanishing gap between the two normal levels when they make closest approach at $A/\omega=2.4$. The enlarged view of the closest approach for the symmetric driving case reveals that there is no true level crossing between Floquet states.

\begin{figure}[htbp]
\center
\includegraphics[width=8cm]{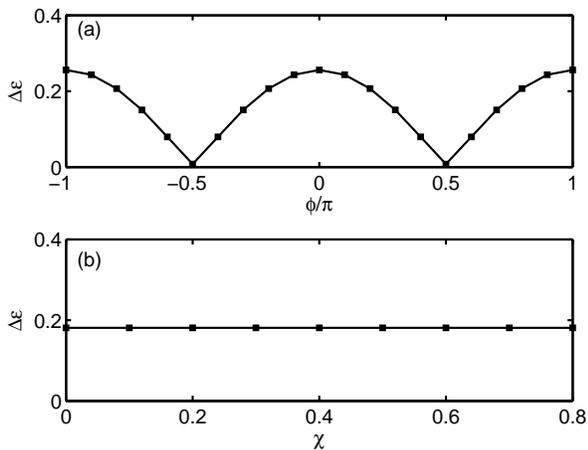}
\caption{(color online) (a) Dependence of $\Delta\varepsilon$  on the phase shift $\phi$. $f=1/4, A/\omega=2.4,\omega=10, v=1,\chi=0.4$.  (b) Dependence of  $\Delta\varepsilon$  on the nonlinearity parameter $\chi$.  $f=1/4, \phi=\pi/4, A/\omega=2.4,\omega=10, v=1$. The quantity $\Delta\varepsilon$ denotes the minimal level spacing between the two normal Floquet states at $A/\omega=2.4$, as indicated in Fig.~\ref{fig1} (b).} \label{fig2}
\end{figure}

In Fig.~\ref{fig2}, we have numerically examined the dependence of the  level spacing $\Delta\varepsilon$ on the phase shift (top panel) and nonlinearity strength (bottom panel). As shown in Fig.~\ref{fig2} (a),  for $\phi=\pm \pi/2$, i.e., in the presence a time-reversal symmetry,   the energy gap $\Delta\varepsilon$ nearly vanishes, and the maximum values of $\Delta\varepsilon$  are reached for $\phi=n\pi$ with $n$ integers (maximally broken time-reversal symmetry, but in the presence of a time-reversal antisymmetry). The dependence of $\Delta\varepsilon$  on the  nonlinearity strength is exemplified in Fig.~\ref{fig2} (a) for a specific case $\phi=\pi/4$. It is clearly seen that $\Delta\varepsilon$ remains unchanged as the  nonlinearity strength varies. The other choice of the phase shift will produce the same result. This means that the spectrum structures of the two normal Floquet states having linear
counterparts are not affected by the presence of  nonlinearity.

\begin{figure}[htbp]
\center
\includegraphics[width=8cm]{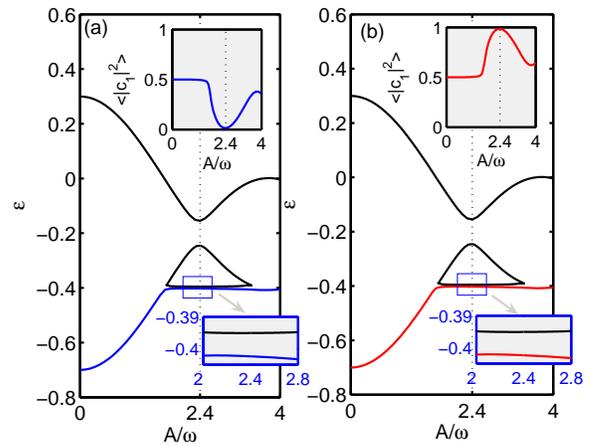}
\caption{(color online) The Floquet state quasienergies versus the driving parameter $A/\omega$ at (a) $\phi=-\pi/4$ and (b) $\phi=\pi/4$. In the two plots, the
top-right insets are the time-averaged population $\langle |c_1|^2\rangle$ for the Floquet state in the lowest quasienergy level, and the bottom-right insets are the enlarged view of the two quasi-degenerate (nearly coincident) quasienergy levels originating from nonlinear bifurcation. The other parameters are $f=1/4,\omega=10, v=1,\chi=0.4$.} \label{fig3}
\end{figure}

Most strikingly, switching the sign of phase shift $\phi$ creates the lowest Floquet
states with opposite population imbalances, as shown in Fig.~\ref{fig3}. By comparing Fig.~\ref{fig3} (a)
and  Fig.~\ref{fig3} (b), we clearly see that both cases of $\phi=\pm \pi/4$ have exactly the same quasienergy spectrum.
For $\phi=\pm \pi/4$ (neither time-reversal symmetric nor time-reversal antisymmetric), there exists a level gap (no level crossing) between the two normal
Floquet states (see the two upper black lines) which have linear analogues, and when the nonlinearity is strong enough in this system [e.g., $\chi=0.4$ in Fig.~\ref{fig3}], there are two nearly coincident (not degenerate) quasienergy levels within a finite interval of
parameter values around $A/\omega=2.4$ (see the bottom insets), which stem from the level bifurcations caused by nonlinearity. In the insets of Fig.~\ref{fig3}, we have also plotted the  cycle-averaged population $\langle |c_1|^2\rangle=\frac{1}{T}\int_0^T|c_1|^2 dt$ for the  Floquet state corresponding to the lowest level.
We may expect that the lowest Floquet state with nearly symmetric population
distribution
continuously evolves into the one with strong population imbalance. As $A/\omega$ is increased from zero to $2.4$, we observe that the variable $\langle |c_1|^2\rangle$
drops down to 0 for $\phi=-\pi/4$, which implies the complete localization at state $|2\rangle$, whereas $\langle |c_1|^2\rangle$ rises up to 1 for $\phi=\pi/4$,
corresponding to complete localization at state $|1\rangle$. Thus, by tuning the phase shift, we can switch between the two strongly localized states with opposite population imbalances.

\section{Physical consequences}
\begin{figure}[htbp]
\center
\includegraphics[width=8cm]{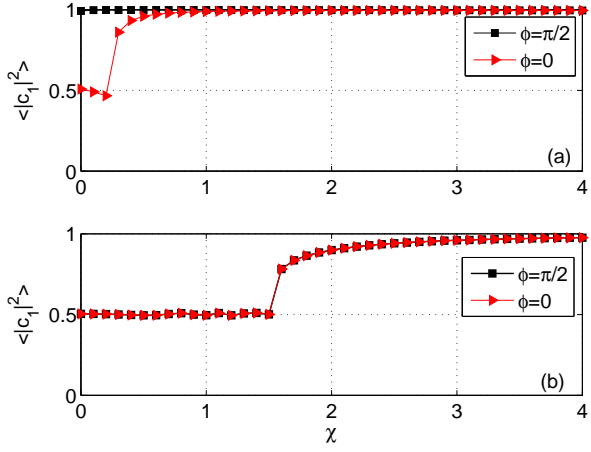}
\caption{(color online) Dependence of the time-averaged population $\langle |c_1|^2\rangle$ on nonlinearity parameter $\chi$ at (a) $A/\omega=2.4$ and (b) $A/\omega=1$. Here the system is initially prepared at state $|1\rangle$, and the average $\langle ...\rangle$ is numerically realized over a long-enough time interval. Two prototypical examples, namely $\phi=0$ (not time-reversal symmetric but time-reversal antisymmetric) and  $\phi=\pi/2$ (time-reversal symmetric) drivings, are compared. Other parameters are the same as in Fig.~\ref{fig1}. } \label{fig4}
\end{figure}

In this section, we will investigate the physical implications of the above-mentioned symmetries on the dynamics of nonlinear two-mode system.
In Fig.~\ref{fig4}, we initialize the system in state $|1\rangle$, and numerically plot the average of the population $|c_1|^2$ over long-enough time interval at $A/\omega=2.4$ (a) and $A/\omega=1$ (b) for $\phi=0$ and $\phi=\pi/2$. By comparison, we find that at $A/\omega=1$, the averages of the population $|c_1|^2$ for $\phi=0$ and $\phi=\pi/2$ exhibit the same (overlapped) dynamics with the same transition to localization for nonlinearity strength above a critical value, while at $A/\omega=2.4$,
 the averages show qualitatively different dynamical features for $\phi=0$ and $\phi=\pi/2$. This can be explained by noting that the Floquet eigenspectra for $\phi=0$ and $\phi=\pi/2$ are the same in the regions away from $A/\omega=2.4$, but they are different in the region around $A/\omega=2.4$.
In Fig.~\ref{fig4} (a), for  $\phi=\pi/2$ (time-reversal symmetric driving), we find that there is no threshold value of nonlinearity for
localization to occur, which corresponds to an almost perfect (but not true) level crossing between the two normal Floquet states at $A/\omega=2.4$.
In this case, we notice the essential role played by the periodic driving on the localization phenomenon, and  in the linear limit, this localization can be connected  to the
well-known CDT phenomenon. In contrast, if the time-reversal symmetry is broken [for example, $\phi=0$  in Fig.~\ref{fig4} (a)], localization occurs
only above a certain critical value of nonlinearity, suggesting the localization being a purely nonlinear phenomenon. This effect is related to the existence of a relatively large level gap ($\Delta \varepsilon$) of the two normal Floquet states when they make closest approach at $A/\omega=2.4$.

\begin{figure}[htbp]
\center
\includegraphics[width=8cm]{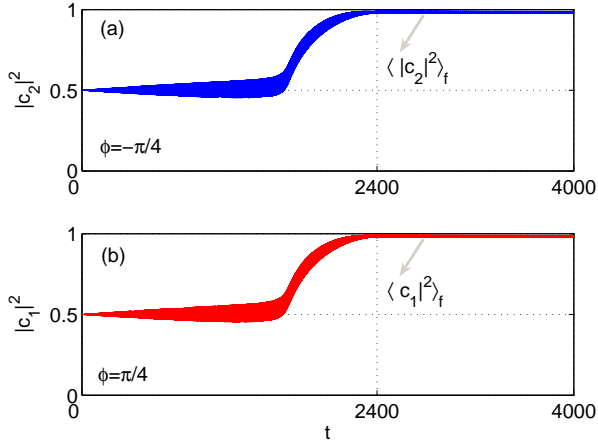}
\caption{(color online) Localization induced by the modulation $S(t)$
with a linearly ramped amplitude $A=\alpha t$, where the dimensionless ramping
rate $\alpha = 0.01$, for the two-mode system \eqref{Dnls1} with two different values of $\phi$. Ramping the modulation amplitude until
$t = 2400$ (amounts to $A/\omega=2.4$) gives the maximum localization at the basis state $|2\rangle$ (blue curve, $\phi=-\pi/4$),  or the other basis state $|1\rangle$ (red curve, $\phi=\pi/4$).  Holding $A$ constant after the certain
time ($t = 2400$, marked by the vertical line) keeps the localization at a
constant level. The system is initialized in its ground state, $c_1(0)=c_2(0)=1/\sqrt{2}$, and the other parameters are the same as in Fig.~\ref{fig3}.} \label{fig5}
\end{figure}

\begin{figure}[htbp]
\center
\includegraphics[width=8cm]{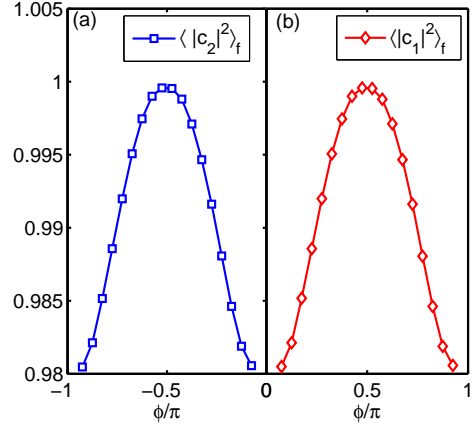}
\caption{(color online) The final localization according to the adiabatic
process outlined in  Fig.~\ref{fig5} as a function of phase shift $\phi$. Here the localization is quantified by the final
time-averaged population of the two modes as $\langle |c_n|^2\rangle_f=\frac{1}{\Delta t}\int_{t_f}^{t_f+\Delta t}|c_n|^2dt$, ($n=1, 2$), where $t_f=2400$ and the averaging time $\Delta t=400$ are used. The initial condition and system parameters are the same as in Fig.~\ref{fig5}.} \label{fig6}
\end{figure}

As noted in the previous section, the harmonic mixing driving field  permits a sensitive control of the population distribution as a function of the phase shift.
For demonstration, we simply ramp the driving amplitude $A$ linearly in time, viz, $A=\alpha t$,  where $\alpha$ is ramping rate.
The system is initialized in its ground state, $c_1(0)=c_2(0)=1/\sqrt{2}$. The driving amplitude is ramped up from zero to $A/\omega=2.4$ for a given $\omega=10$, then it is held
constant hereafter.  When the ramping rate takes a low value $\alpha=0.01$, we expect the system
to adiabatically follow the lowest Floquet state. Two different scenarios of localization dynamics are identified in Fig.~\ref{fig5}, for two different values of the
phase shift $\phi$. For $\phi=-\pi/4$, the time-evolving state is finally localized at $|2\rangle$ and stays there subsequently.
When the phase shift is changed to a positive value, $\phi=\pi/4$,
the time-evolving state becomes concentrated in  the other basis state $|1\rangle$. Thus, states with opposite population
imbalances can be selectively targeted, which may help to control localization process. To provide a more
complete analysis on this phase control of population imbalance, we evaluate the final
time-averaged population of the two modes as $\langle |c_n|^2\rangle_f=\frac{1}{\Delta t}\int_{t_f}^{t_f+\Delta t}|c_n|^2dt$, ($n=1, 2$), where
$t_f$ represents the time instant when the linearly-ramping
modulation amplitude reaches the value $A/\omega=2.4$ (giving the maximum  localization),
and $\Delta t$ represents the long-enough averaging time interval
(during which the modulation amplitude stays constant). As expected, the degree of final localization is symmetric with respect to $\phi=0$ [see Fig.~\ref{fig6}], and which one of the two basis states is highly occupied depends on the sign of the phase shift.

Now it is natural to put a simple question: why can the change of the phase shift lead to inversion of population imbalance?  At present, the analytical results for
the question is still missing in the  literatures, and thus this problem calls for analytical insights.

\section{Perturbative analysis}
To gain analytical insight, in this section we perform a multiple-scale asymptotic analysis of the model \eqref{Dnls1} in the high frequency
limit $\omega\gg \{v, \chi\}$(see, for instance, Refs.~\cite{Longhi2012, Zhou2013, Luo2014, Luo2021}). For this purpose, we first introduce the slowly varying functions $a_1$ and $a_2$ through the transformation
\begin{equation}\label{Forma1}
  c_1=a_1e^{-i\int\frac{S(t)}{2}dt},~
  c_2=a_2e^{i\int\frac{S(t)}{2}dt}.
\end{equation}
Substituting the  transformation \eqref{Forma1} into Eq.~\eqref{Dnls1}, we obtain the following coupled equation
\begin{align}\label{Dnls2}
  i\frac{da_1}{dt}= &-\frac{v'}{2}a_2-\chi|a_1|^2a_1,\nonumber\\
  i\frac{da_2}{dt}= &-\frac{v'^*}{2}a_1-\chi|a_2|^2a_2,
\end{align}
where $v'=ve^{i\int{S(t)}dt}=v\exp{[i\frac{A}{\omega}\cos\omega t+i\frac{Af}{2\omega}\cos(2\omega t+\phi)]}$.

Denoting
\begin{equation}
F(t)=\exp{[i\frac{A}{\omega}\cos\omega t+i\frac{Af}{2\omega}\cos(2\omega t+\phi)]},
\end{equation}
and introducing a new variable
\begin{equation}\label{new var}
  \tau=\omega t,~\epsilon=\frac{v}{\omega},
\end{equation}
then Eq.~\eqref{Dnls2} reads as
\begin{align}\label{Dnls3}
  i\frac{da_1}{d\tau}= &-\frac{\epsilon}{2}F(\tau)a_2-\epsilon\frac{\chi}{v}|a_1|^2a_1,\nonumber\\
  i\frac{da_2}{d\tau}= &-\frac{\epsilon}{2}F^*(\tau)a_1-\epsilon\frac{\chi}{v}|a_2|^2a_2.
\end{align}

 Let us look for a solution to Eq.~\eqref{Dnls3} as a power-series expansion in the smallness parameter $\epsilon$:
\begin{equation}\label{expand a}
  a_j=a_j^{(0)}+\epsilon a_j^{(1)}+\epsilon^2 a_j^{(2)}+\cdots,~j=1,2,
\end{equation}
and introduce multiple time scales $T_0=\tau,~T_1=\epsilon\tau,~T_2=\epsilon^2\tau,\dots.$

By using the derivative rule
$\frac{d}{d\tau}=\frac{\partial}{\partial T_0}+\epsilon\frac{\partial}{\partial T_1}+
  \epsilon^2\frac{\partial}{\partial T_2}+\cdots,$ and the fact
  \begin{equation*}
  |a_j|^2a_j=|a_j^{(0)}|^2a_j^{(0)}+\epsilon[2|a_j^{(0)}|^2a_j^{(1)}+(a_j^{(0)})^2a_j^{*(1)}]+\cdots,~j=1,2,
\end{equation*}
and substituting Eq.~\eqref{expand a} into Eq.~\eqref{Dnls3}, we obtain a hierarchy of equations for successive corrections to
$a_{1,2}$ at the various orders in $\epsilon$.
At the leading order $\epsilon^0$, we find
\begin{equation}\label{0order}
  i\partial_{T_0} a_j^{(0)}=0, ~~~a_j^{(0)}=A_j(T_1,T_2,\cdots),~j=1,2,
\end{equation}
where the amplitudes $A_{1,2}(T_1,T_2,\cdots)$ are functions of the slow time variables
$T_1,T_2,\cdots$, but independent of the fast time variable $T_0$. At order $\epsilon^1$ one has
\begin{align}\label{1order}
  i\partial_{T_1} a_1^{(0)}+i\partial_{T_0} a_1^{(1)}=&-\frac{F(\tau)}{2}a_2^{(0)}-\frac{\chi}{v}|a_1^{(0)}|^2a_1^{(0)}\nonumber\\
  i\partial_{T_1} a_2^{(0)}+i\partial_{T_0} a_2^{(1)}=&-\frac{F^*(\tau)}{2}a_1^{(0)}-\frac{\chi}{v}|a_2^{(0)}|^2a_2^{(0)}.
\end{align}
For the convenience of our discussion, we simplify equation \eqref{1order} as
\begin{align}\label{1orders}
 i\partial_{T_1} A_1+i\partial_{T_0} a_1^{(1)}=&-\frac{F(\tau)}{2}A_2-\frac{\chi}{v}|A_1|^2A_1\nonumber\\
  i\partial_{T_1} A_2+i\partial_{T_0} a_2^{(1)}=&-\frac{F^*(\tau)}{2}A_1-\frac{\chi}{v}|A_2|^2A_2.
\end{align}
To avoid the occurrence of secularly growing terms in the solutions $a_1^{(1)}$ and $a_2^{(1)}$, the solvability conditions
\begin{equation}\label{solvcondi1}
  i\partial_{T_1} A_1=-\frac{\overline{F(\tau)}}{2}A_2-\frac{\chi}{v}|A_1|^2A_1,
  ~~i\partial_{T_1} A_2=-\frac{\overline{F^*(\tau)}}{2}A_1-\frac{\chi}{v}|A_2|^2A_2
\end{equation}
must be satisfied. Throughout our paper, the overline denotes the time average with respect to the fast time variable $T_0$.
$F(\tau)$ can be expanded by using of the first kind Bessel function $J_{\alpha}(x)$ with order $\alpha$,
\begin{equation*}
  F(\tau)=\sum_{m,n}J_n(\frac{A}{\omega})J_m(\frac{Af}{2\omega})i^{m+n}e^{i(2m+n)\tau}e^{im\phi},
\end{equation*}
which gives
\begin{equation}\label{F aver}
  \overline{F(\tau)}=\sum_{m}J_{-2m}(\frac{A}{\omega})J_m(\frac{Af}{2\omega})i^{-m}e^{im\phi}.
\end{equation}

It follows from Eqs.~\eqref{1orders}, \eqref{solvcondi1} and \eqref{F aver}, that the amplitudes $a_{1,2}^{(1)}$
at order $\epsilon$ are given by
\begin{align}\label{a11 a21}
  a_1^{(1)}&=-i\int(-\frac{F(\tau)}{2}A_2+\frac{\overline{F(\tau)}}{2}A_2)d\tau=A_2\Phi(\tau),\nonumber\\
  a_2^{(1)}&=-i\int(-\frac{F^*(\tau)}{2}A_2+\frac{\overline{F^*(\tau)}}{2}A_2)d\tau=-A_1\Phi^*(\tau),
\end{align}
where $\Phi(\tau)=\sum_{n\neq-2m}\frac{1}{2m+n}J_{n}(\frac{A}{\omega})J_m(\frac{Af}{2\omega})i^{m+n}e^{im\phi}e^{i(2m+n)\tau}$.

At the next order $\epsilon^2$, we have
\begin{align}\label{2order}
i\left(\partial_{T_2} a_1^{(0)}+\partial_{T_1} a_1^{(1)}+\partial_{T_0} a_1^{(2)}\right)
 &=-\frac{F(\tau)}{2}a_2^{(1)}-\frac{\chi}{v}[2|a_1^{(0)}|^2a_1^{(1)}\nonumber\\
 &+(a_1^{(0)})^2a_1^{*(1)}],\nonumber\\
i\left(\partial_{T_2} a_2^{(0)}+\partial_{T_1} a_2^{(1)}+\partial_{T_0} a_2^{(2)}\right)
 &=-\frac{F^*(\tau)}{2}a_1^{(1)}-\frac{\chi}{v}[2|a_2^{(0)}|^2a_2^{(1)}\nonumber\\
 &+(a_2^{(0)})^2a_2^{*(1)}].
\end{align}

In order to avoid the occurrence of secularly growing terms in the solutions $a_1^{(2)}$ and $a_2^{(2)}$,
the following solvability conditions must be satisfied:
\begin{align}\label{solvcondi3}
  i\partial_{T_2} A_1=&-\overline{\frac{F(\tau)}{2}a_2^{(1)}}=A_1\overline{\frac{F(\tau)\Phi^*(\tau)}{2}}=\frac{\delta}{2}A_1,\nonumber\\
  i\partial_{T_2} A_2=&-\overline{\frac{F(\tau)}{2}a_1^{(1)}}=-A_2\overline{\frac{F^*(\tau)\Phi(\tau)}{2}}=-\frac{\delta^*}{2}A_2,
\end{align}
 where
 \begin{align}\label{delta}
   \delta =\overline{F(\tau)\Phi^*(\tau)}&=\sum_{m,M,l,M\neq0}\frac{(-1)^{M-l}}{M}i^{2M-m-l}e^{i(m-l)\phi}J_{M-2m}(\frac{A}{\omega})\nonumber\\
   &\times J_{M-2l}(\frac{A}{\omega})J_{m}(\frac{Af}{2\omega})J_{l}(\frac{Af}{2\omega}).
 \end{align}
 Thus the evolution of the amplitudes $A_{1,2}$ up to the second-order long time scale is given by
 \begin{equation}\label{AjSecond}
   i\frac{dA_j}{d\tau}=i\epsilon\partial_{T_1} A_j+i\epsilon^2\partial_{T_2} A_j,~~j=1,2.
 \end{equation}
 Substituting equations \eqref{0order}, \eqref{solvcondi1} and \eqref{solvcondi3} into equation \eqref{AjSecond}, we obtain
 \begin{align}\label{scale Aj equ}
  i\frac{dA_1}{d\tau}&=-\epsilon\frac{\overline{F(\tau)}}{2}A_2-\epsilon\frac{\chi}{v}|A_1|^2A_1+\epsilon^2\frac{\delta}{2}A_1,
  \nonumber\\
  i\frac{dA_2}{d\tau}&=-\epsilon\frac{\overline{F^*(\tau)}}{2}A_1-\epsilon\frac{\chi}{v}|A_2|^2A_2-\epsilon^2\frac{\delta^*}{2}A_2.
\end{align}
 Substituting \eqref{new var} into \eqref{scale Aj equ}, we have
 \begin{align}\label{Ajdiff}
  i\frac{dA_1}{dt}&=-v\frac{\overline{F(\tau)}}{2}A_2-\chi|A_1|^2A_1+\frac{v^2}{\omega}\frac{\delta}{2}A_1,\nonumber\\
  i\frac{dA_2}{dt}&=-v\frac{\overline{F^*(\tau)}}{2}A_1-\chi|A_2|^2A_2-\frac{v^2}{\omega}\frac{\delta^*}{2}A_2.
\end{align}

 Let $v'=v\overline{F(\tau)}$ and $\delta'=\frac{v^2}{\omega}\delta$, then Eq.~\eqref{Ajdiff} reads
 \begin{align}\label{Ajdiff2}
  i\frac{dA_1}{dt}&=\frac{\delta'}{2}A_1-\chi|A_1|^2A_1-\frac{v'}{2}A_2,\nonumber\\
  i\frac{dA_2}{dt}&=-\frac{\delta'^*}{2}A_2-\chi|A_2|^2A_2-\frac{v'^{\ast}}{2}A_1.
\end{align}
Numerical investigations (not shown here) reveal that from Eq.~\eqref{Ajdiff2} (which is accurate up to the time scale $\sim 1/\epsilon^2$), we  can  recover all the known quasienergy spectrums based on the original model \eqref{Dnls1}.  Thus, the effective equation \eqref{Ajdiff2}  constitutes a strong  analytical basis for understanding the dynamical features of the original system with different time-space symmetries. It
is noteworthy that, in the effective equation \eqref{Ajdiff2}, a second-order static dc-bias $\delta'$ appears as a surprise, apart from the
coupling strength $v$ replaced by $v'$. In the following, we will rigorously prove
that the effective tunneling rate $v'$ and the second-order detuning $\delta'$ enjoy some very
interesting properties.

(i) $v'$ is a real number when $\phi=\pm\pi/2$ [$\phi\in [-\pi,\pi)$], and is a complex number otherwise.

We write down
 \begin{equation}\label{F aver1}
  \overline{F(\tau)}=\sum_{m}J_{-2m}(\frac{A}{\omega})J_m(\frac{Af}{2\omega})i^{-m}e^{im\phi},
\end{equation}
and its complex conjunction
\begin{equation}\label{F aver2}
  \overline{F(\tau)}^*=\sum_{m}J_{-2m}(\frac{A}{\omega})J_m(\frac{Af}{2\omega})({-i})^{-m}e^{-im\phi}.
 \end{equation}
 Subtracting  \eqref{F aver2} from \eqref{F aver1} yields
\begin{align}\label{F aver3}
  \overline{F(\tau)}-\overline{F(\tau)}^*=&\sum_{m}J_{-2m}(\frac{A}{\omega})J_m(\frac{Af}{2\omega})[e^{im(\phi-\frac{\pi}{2})}-e^{-im(\phi-\frac{\pi}{2})}]
  \nonumber\\
=&2i\sum_{m}J_{-2m}(\frac{A}{\omega})J_m(\frac{Af}{2\omega})\sin\big[m(\phi-\frac{\pi}{2})\big ].
\end{align}

 Evidently, when $\phi=\pm\pi/2$ [$\phi\in [-\pi, \pi)$], $\overline{F(\tau)}=\overline{F(\tau)}^*$, thus $v'=v\overline{F(\tau)}$ is a real number.
 If otherwise, i.e., when $\phi\neq\pm\pi/2$ [$\phi\in [-\pi, \pi)$], $\overline{F(\tau)}$ and $v'=v\overline{F(\tau)}$ are, in general, complex numbers.

(ii) $\delta'$ is always a real number.

 By using the expression \eqref{delta} of $\delta$, one can obtain that
\begin{align}\label{delta conj}
   \delta^*=\sum_{m,l,M,M\neq0} &\frac{(-1)^{M-l}}{M}(-i)^{2M-m-l}e^{-i(m-l)\phi}J_{M-2m}(\frac{A}{\omega})
   J_{M-2l}(\frac{A}{\omega})\nonumber\\
   &\times J_{m}(\frac{Af}{2\omega})J_{l}(\frac{Af}{2\omega})\nonumber\\
   =\sum_{m,l,M,M\neq0} &\frac{(-1)^{M-m}}{M}(-i)^{2M-m-l}e^{-i(l-m)\phi}J_{M-2m}(\frac{A}{\omega})
   J_{M-2l}(\frac{A}{\omega})\nonumber\\
   &\times J_{m}(\frac{Af}{2\omega})J_{l}(\frac{Af}{2\omega}).
 \end{align}
Here we have made the exchange $m\leftrightarrow l$. Since $(-1)^{M-m}(-1)^{2M-m-l}=(-1)^{M-l}$, we get from Eq.~\eqref{delta conj}
that $\delta=\delta^*$, thus $\delta$ is a real number.

(iii) $\delta(-\phi)=-\delta(\phi)$, and $\delta$ must be zero when $\phi=0$ [i.e., when $S(t)$ possesses the antisymmetry $S(t)=-S(-t)$].

Separating $M>0$ and $M<0$ parts of the expression \eqref{delta} of $\delta$, making the transformation $M\rightarrow-M$ for negative $M$,
we obtain
\begin{align}\label{delta other1}
   \delta=\sum_{m,M,l,M>0} &\frac{(-1)^{M-l}}{M}i^{2M-m-l}e^{i(m-l)\phi}J_{M-2m}(\frac{A}{\omega})
   J_{M-2l}(\frac{A}{\omega})\nonumber\\
   &\times J_{m}(\frac{Af}{2\omega})J_{l}(\frac{Af}{2\omega}) \nonumber\\
   -\sum_{m,M,l,M>0} &\frac{(-1)^{-M-l}}{M}i^{-2M-m-l}e^{i(m-l)\phi}J_{-M-2m}(\frac{A}{\omega})\nonumber\\
     &\times J_{-M-2l}(\frac{A}{\omega})J_{m}(\frac{Af}{2\omega})J_{l}(\frac{Af}{2\omega}).
 \end{align}
Changing the  summation indices $m,l$ into $-m,-l$ in the second summation in \eqref{delta other1}, we have
\begin{align}\label{delta other2}
   \delta=\sum_{m,M,l,M>0} &\frac{(-1)^{M-l}}{M}i^{2M-m-l}e^{i(m-l)\phi}J_{M-2m}(\frac{A}{\omega})
   J_{M-2l}(\frac{A}{\omega}) \nonumber\\
     &\times J_{m}(\frac{Af}{2\omega})J_{l}(\frac{Af}{2\omega}) \nonumber\\
   -\sum_{m,M,l,M>0} &\frac{(-1)^{-M+l}}{M}i^{-2M+m+l}e^{-i(m-l)\phi}J_{-M+2m}(\frac{A}{\omega}) \nonumber\\
     &\times J_{-M+2l}(\frac{A}{\omega})J_{-m}(\frac{Af}{2\omega})J_{-l}(\frac{Af}{2\omega}).
 \end{align}
By using  the relation of Bessel function $J_{-\alpha}=(-1)^{\alpha}J_{\alpha}$ and the fact that $(-1)^{-m-l}=i^{-2m-2l}$,
then we obtain the alternative form of $\delta$ from Eq.~\eqref{delta other2}
 \begin{align}\label{delta other}
   \delta=\sum_{m,M,l,M>0} &\frac{(-1)^{M-l}}{M}i^{2M-m-l}(e^{i(m-l)\phi}-e^{-i(m-l)\phi})J_{M-2m}(\frac{A}{\omega}) \nonumber\\
  &\times J_{M-2l}(\frac{A}{\omega})J_{m}(\frac{Af}{2\omega})J_{l}(\frac{Af}{2\omega}).
 \end{align}
 When $\phi=0$, it follows from Eq.~\eqref{delta other} that $\delta=0$. It is easy to see that $\delta(-\phi)=-\delta(\phi)$
  from Eq.~\eqref{delta other}.

\begin{figure}[htbp]
\center
\includegraphics[width=8cm]{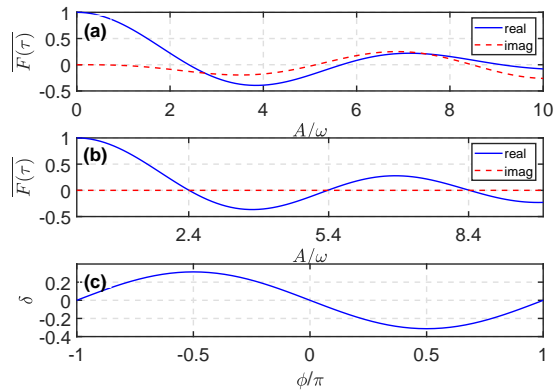}
\caption{(color online) $\overline{F(\tau)}$ (the renormalized factor of coupling strength)  as a function of driving parameter $A/\omega$ for (a) $\phi=0$ (breaking the temporal symmetry) and (b) $\phi=\pi/2$ (symmetric driving). The blue solid line and red dashed line denote the real and imaginary parts of $\overline{F(\tau)}$ respectively. (c) $\delta$ (describing the effective bias) as a function of phase shift $\phi$ at a particular driving parameter  $A/\omega=2.4$. The other parameters are $f=1/4,\omega=10, v=1,\chi=0.4$. } \label{fig7}
\end{figure}

To corroborate these analytical results, we numerically calculate the two quantities  $\overline{F(\tau)}$ and $\delta$ as shown in Fig.~\ref{fig7}. The dependencies of $\overline{F(\tau)}$ on the driving parameter $A/\omega$ are illustrated in Figs.~\ref{fig7} (a) and (b) for $\phi=0$ and $\phi=\pi/2$ respectively, which verifies the fact that $\overline{F(\tau)}$ [hence the effective coupling strength $v'=v\overline{F(\tau)}$] is a real number when $\phi=\pi/2$, and is generally a complex number when $\phi=0$. Note that, if the time-reversal symmetry ($\phi=\pi/2$) is preserved,  $\overline{F(\tau)}$ (consequently, the effective coupling strength) vanishes identically at some certain driving parameter values $A/\omega=2.4, 5.4, 8.4$. Whereas if the time-reversal symmetry is violated [see, e.g., $\phi=0$ in Fig.~\ref{fig7}(a)], $\overline{F(\tau)}$ is not equal to zero for any value of $A/\omega$.  As shown in Fig.~\ref{fig7} (c), we also obtain a dependence of the effective detuning $\delta$ on $\phi$ with sign changes, as expected. We clearly observe that the effective detuning $\delta$ takes maximum (minimum) values at $\phi=\pm \pi/2$ (harmonic mixing signal is symmetric but not antisymmetric), but instead vanishes at $\phi=0$ (harmonic mixing signal is antisymmetric).

 The properties for  $\overline{F(\tau)}$ and $\delta$ have the following physical implications. First, when $\phi=\pi/2$ such that the temporal symmetry $S(t+t_0)=S(-t+t_0)$ is preserved, $\overline{F(\tau)}$  vanishes at  $A/\omega=2.4$, while $\delta$ takes nonzero value. The eigenvalues of Eq.~\eqref{Ajdiff2} with an effectively undriven
(time-averaged) Hamiltonian
are the quasienergies of  the original time-dependent
quantum system \eqref{Dnls1}.
 In the linear limit, from \eqref{Ajdiff2} we obtain the eigenvalues as $E_{\pm}=\pm\frac{1}{2}\sqrt{\delta'^2+|v\overline{F(\tau)}|^2}$. When the driving parameter is chosen as $A/\omega=2.4$ such that $\overline{F(\tau)}$  vanishes,  there will be a minimum of the level spacing,
which is fixed by an extremely small value $\Delta\varepsilon=\delta'$, and confirms that the quasienergies of the two normal Floquet states (having linear counterparts) form an anticrossing rather than a crossing. Thus, the common
explanation of CDT fails. Here, the CDT for a  time-reversal symmetric system comes from vanishing of the effective coupling strength, not from the quasienergy degeneracy. Second,  when $\phi\neq\pi/2$ such that the time-reversal symmetry $S(t+t_0)=S(-t+t_0)$ is broken, $\overline{F(\tau)}$ (hence the  effective coupling strength)  does not vanish for any value of  $A/\omega$. As illustrated in Fig.~\ref{fig2} (b), the energy gap $\Delta\varepsilon$ (the minimum level spacing between two normal Floquet states) shows roughly no dependence on nonlinearity, and its value can be approximated (or accurately given) by $\Delta\varepsilon=\sqrt{\delta'^2+|v\overline{F(\tau)}|^2}$ at $A/\omega=2.4$, where $|\overline{F(\tau)}|$ takes minimum (nonzero) value. This  approximation is made in the case $\delta'\neq 0$ (making the two normal Floquet states a little bit unbalanced), where we neglect the negligibly small nonlinear energy offset, i.e., the term $\chi(|A_1|^2-|A_2|^2)$. Note that here the minimum value of $|\overline{F(\tau)}|$ is  nonzero and much larger than the absolute value of second-order bias $|\delta'|$. Thus, the broken  time-reversal symmetry leads to a relatively large energy gap $\Delta\varepsilon$ between the two normal Floquet states having linear counterparts. In this case, level bifurcation (resulting in the emergence of new localized nonlinear Floquet states) and suppression of tunneling occur only for  nonlinearity beyond a certain threshold value. Third, the change of the phase shift can create the lowest Floquet states with opposite population imbalances. This can be reasoned as follows. When $\phi\neq0$ [$S(t)\neq-S(-t)$], a nonzero second-order bias $\delta'=\frac{v^2}{\omega}\delta$ is generated by the broken time-reversal antisymmetry. When $\delta'\neq0$, under inversion $\delta'\rightarrow-\delta'$, we conclude that for any eigenstate  $(A_1', A_2')^T$ to \eqref{Ajdiff2},
 there is a partner eigenstate $(A_2', A_1')^T$ of \eqref{Ajdiff2} with $\delta'$ replaced by $-\delta'$ at the same energy.
 This implies  inversion $\delta'\rightarrow-\delta'$ does not change the  energy spectrum, but flip the sign of population imbalance
 corresponding to the same energy. Because of $\delta(-\phi)=-\delta(\phi)$, the inversion of population imbalance can be expected by changing the phase shift $\phi$.

\section{Conclusion}
In summary, the symmetry and underlying physics of the nonlinear two-state system
driven by a harmonic mixing field have
been studied analytically and numerically.  A multiple-scale
asymptotic analysis is used to understand the essential  physics with  different time-space symmetries.
By use of the effective description
valid up to the second order of $1/\omega$, we have clarified the origin of the CDT in the time-reversal symmetric
two-state system, and explained the reason why the broken time-reversal antisymmetry can induce the inversion of population imbalance between two modes. These analytical results establish an intimate connection between
symmetry breakings and the relevance of dynamical properties in the nonlinear two-mode system. Although various  aspects of the nonlinear two-mode system have been explored previously,
the analytical results on connection between symmetry and dynamical features of the system have not been addressed before and the present paper
fills the gap in literatures.

\acknowledgments
The work was supported by the Natural Science Foundation of Zhejiang Province, China (Grant No. LY21A050002), the National
Natural Science Foundation of China (Grant No. 11975110), the Scientific and Technological Research Fund of
Jiangxi Provincial Education Department (Grant No. GJJ211026), and Zhejiang Sci-Tech University Scientific Research
Start-up Fund (Grant No. 20062318-Y), the Scientific Research Foundation of Hunan Provincial Education Department (Grant No. 21B0063), and the Hunan Provincial Natural Science Foundation of China (Grant No. 2021JJ30435).

Xianchao Le and Zhao-Yun Zeng contributed equally.

\end{document}